\documentclass[aps,twocolumn,superscriptaddress,floatfix,noeprint]{revtex4-2}
\usepackage{algorithm}
\usepackage{algpseudocode}
\usepackage{blindtext}
\usepackage{bm}
\usepackage{braket}
\usepackage{hyperref}
\usepackage{amssymb}
\usepackage{amsmath}
\usepackage{graphicx}
\usepackage{float}
\usepackage{bbold}
\usepackage{xcolor}
\usepackage{mathtools}
\algrenewcommand\algorithmicrequire{\textbf{Input:}}
\algrenewcommand\algorithmicensure{\textbf{Output:}}
\newcommand{\I}{\mathrm{i}}
\newcommand{\id}{\mathbb{1}}
\newcommand{\vect}[1]{\ensuremath{\mathbf{#1}}}

\renewcommand{\t}[1]{\textrm{#1}}
\begin{document}
\title{Using adaptiveness and causal superpositions against noise in quantum metrology}
\author{Stanis{\l}aw Kurdzia{\l}ek}
\thanks{These two authors contributed equally to the project.}
\affiliation{Faculty of Physics, University of Warsaw, Pasteura 5, 02-093 Warszawa, Poland}
\author{Wojciech G\'orecki}
\thanks{These two authors contributed equally to the project.}
\affiliation{Faculty of Physics, University of Warsaw, Pasteura 5, 02-093 Warszawa, Poland}
\author{Francesco Albarelli}
\affiliation{Dipartimento di Fisica ``Aldo Pontremoli'', Università degli Studi di Milano, via Celoria 16, 20133 Milan, Italy}
\affiliation{Istituto Nazionale di Fisica Nucleare, Sezione di Milano, via Celoria 16, 20133 Milan, Italy}
\author{Rafa{\l}  Demkowicz-Dobrza{\'n}ski} 
\affiliation{Faculty of Physics, University of Warsaw, Pasteura 5, 02-093 Warszawa, Poland}

\begin{abstract}
We derive new bounds on achievable precision in the most general adaptive quantum metrological scenarios.
The bounds are proven to be asymptotically saturable and equivalent to the known parallel scheme bounds in the limit of large number of channel uses.
This completely solves a long standing conjecture in the field of quantum metrology on the asymptotic equivalence between parallel and adaptive strategies.
The new bounds also allow to easily assess the potential benefits of invoking non-standard causal superposition strategies, for which we prove, similarly to the adaptive case, the lack of asymptotic advantage over the parallel ones.
\end{abstract}

\maketitle

\begin{figure}
\includegraphics[width=0.95\columnwidth]{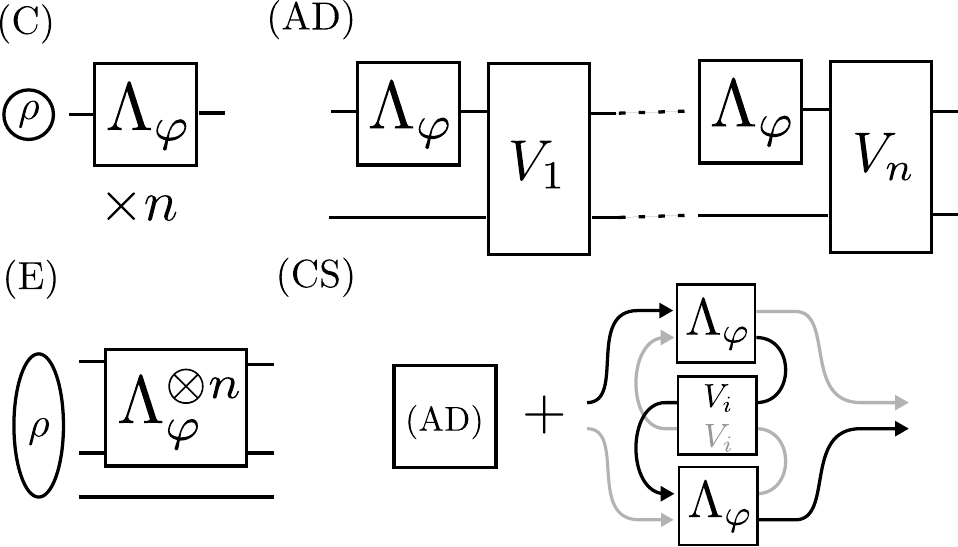}
\caption{Metrological schemes utilizing ``four levels of quantumness'': (C) channels probed independently (basic use of quantum coherence); (E) channels probed in parallel using a general entangled state, with ancillary systems potentially involved; (AD)
general adaptive (active quantum feedback) strategies; (CS) causal superposition strategies, where additionally channels may be probed in a superposition of different causal orders.}
\label{fig:schemes}
\end{figure}

\paragraph*{Introduction.}
In the field of quantum information and quantum technologies, 
one can distinguish three levels of \emph{quantumness}
that are behind the boost in performance of various communication \cite{Gisin2007, Feihu2020}, computational \cite{Preskill2018} or metrological tasks \cite{Giovannetti2011, Degen2017, Pirandola2018}. The most rudimentary one is \emph{quantum coherence} (C), which refers to the potential of having a single quantum system in the state of quantum superposition. This is already enough to implement secure quantum key distribution protocols \cite{Bennett1984} or even reach the Heisenberg limit in noiseless quantum metrology, provided a given quantum probe can pass through a sensing channel multiple times \cite{Higgins2007, Braun2018}.
The next level is \emph{entanglement} (E), where quantum coherence present in multi-partite systems manifests itself
in the form of non-classical correlations. This quantumness level is crucial to guarantee quantum speed-up in computational tasks \cite{Jozsa2003} as well as to assure the ultimate security in the so-called device-independent quantum key distribution \cite{Acin2007}. In quantum metrology, it had long been appreciated as the way to boost the precision in optical and atomic interferometric tasks \cite{Huelga1997, Pezze2009, Demkowicz-Dobrzanski2015a, Dowling2015, Pezze2018}, either in the form
of somehow overhyped N00N states \cite{Bollinger1996, Dowling2008} or much more practical optical and atomic squeezed states \cite{Caves1981, Schnabel2016}. Finally,
exploiting the quantum potential to its limits, one can consider \emph{adaptive} (AD) or \emph{active quantum feedback}  strategies, where the probes are entangled with noiseless ancillary systems, 
and quantum control operations may actively modify the probe system that will be sent to the subsequent channel based on the information obtained so far \cite{Demkowicz-Dobrzanski2014, Demkowicz-Dobrzanski2017, PengAdaptiveControTimeDependentHamiltonians2017,Pirandola2017, Wan2022}, 
see Fig.~\ref{fig:schemes}. Such protocols represent the most general channel sensing schemes, containing (E) as a special case and  encompassing in particular all quantum error-correcting strategies widely used in the whole field of quantum information processing to counter noise \cite{Kribs2005, Terhal2015,Zhou2017,Layden2019}. 

Interestingly, in the absence of noise, (AD) strategies 
provide no advantage over optimal (E) strategies \cite{Giovaennetti2006}. In the presence of noise, however, some advantages have been observed in the small-number-of-uses regime where a direct search of optimal metrological protocols could be carried out \cite{Demkowicz-Dobrzanski2014,Sekatski2016,Yang2019,Altherr2021,Pereira2021,Liu2022f}.
In 2014 a conjecture has been formulated \cite{Demkowicz-Dobrzanski2014} predicting no asymptotic advantage of (AD) over (E). A notable progress in answering this fundamental question has been made in 2021 \cite{Zhou2019e,Zhou2020}, when it was demonstrated that in the models where quantum coherence cannot be protected against noise on arbitrary scale, and hence the Heisenberg scaling (HS) is not achievable, (AD) strategies offer no asymptotic advantage over (E).
Still, the full answer to the question was lacking, mainly due to the fact that the bounds used there were not tight enough.

In this paper, utilizing our new bounds, we indeed answer the conjecture in an affirmative way, proving in full generality that (AD) strategies provide no asymptotic advantage over (E). As negative as it may sound, the result by no means implies that (AD) strategies are useless. In fact, our bounds allow to clearly pinpoint the potential advantage one may expect in the finite number-of-uses regime, and easily observe how the advantage fades away when approaching the asymptotic limit of large number of channel uses. On a more practical side, adaptive strategies may sometimes be in fact easier to implement than parallel, as they may not necessarily require entangling large number of particles, while obtaining the same effect via small scale entanglement + active feedback.

Even though the ``three levels of quantumness'' listed above appear to cover all quantum aspects of metrological protocols, an intriguing idea was put forward of considering 
\emph{causal superposition strategies} (CS) where different channels are being probed in a  superposition of different causal orders \cite{Araujo2014, Mukhopadhyay2018a,Frey2019a,Zhao2019a, Chapeau-Blondeau2021, Liu2022f,Wechs2021,Bavaresco2021}. Advantages of such a strategy over the most general (AD) strategy have been observed, but no efficiently computable bounds have been proposed. In this paper, we provide bounds valid also for this more general class of protocols and show their asymptotic equivalence to (AD) and (E), which also means that (CS) strategies cannot surpass the HS~\footnote{A (CS) strategy was shown to achieve super-Heisenberg scaling in a quantum metrology problem with infinite-dimensional systems~\cite{Zhao2019a}.
This does not contradict our results, since in this work we derive bounds for finite-dimensional systems.}.

%
%
%
%
%

\paragraph*{Introductory example.}
Let us start with the most elementary yet very illuminating example of a noisy metrological model, where it is possible to remove noise while assuring the preservation of HS of precision in the asymptotic regime.
Consider a single qubit channel $\Lambda_\varphi(\cdot) = \sum_k K_{\varphi,k} \cdot K_{\varphi,k}^{\dagger}$, where $K_{\varphi,k} = U_{\varphi} K_k$, 
\begin{equation}
\label{eq:krausdephaseperp}
U_{\varphi} = e^{- \frac{i}{2}\sigma_z \varphi}, \quad K_{1} = \sqrt{p} \openone, \quad K_{2} = \sqrt{1-p} \sigma_x.
\end{equation}
The channel represents dephasing of a qubit along the $x$ axis of the Bloch ball (the operator $K_2$ may be understood as a $\sigma_x$ error occurring with probability $1-p$) and the subsequent rotation $U_\varphi$ of the state around the $z$ axis by angle $\varphi$, where $\varphi$ is the parameter to be estimated---a similar model has been used in an experimental demonstration of quantum error-correction enhanced metrology in NV-center sensing setups~\cite{Unden2016}, as well as in \cite{Brask2013} where the possibility of beating the standard scaling (SS) in presence of transversal noise was shown.
In case of a single channel use, $n=1$, the effect of noise may be completely mitigated by choosing the input state as
$\ket{\psi^{(1)}}=\ket{+}=(\ket{0}+\ket{1})/\sqrt{2}$. This state is not affected by $\sigma_x$ error and the output state
$\ket{\psi_\varphi} = (\ket{0} + e^{i \varphi} \ket{1})/\sqrt{2}$
represents a noiseless phase encoding.
We will quantify the performance of a given protocol using the \emph{quantum Fisher information} (QFI) \cite{Helstrom1976,Braunstein1994} of the output state, which in this case is  $F^{(1)}=1$ (we recall the definition of the QFI in Appendix~\ref{app:example}).

Assume now that we can use the channel twice, $n=2$. If we send the optimal single qubit probes independently to each of the channels, we get the QFI value $F^{(2)}_{\t{C}} = 2$. 
We can, however, also consider a parallel strategy involving an entangled input state $\ket{\psi^{(2)}}= (\ket{00} + \ket{11})/\sqrt{2}$ (the N00N state). 
In this case if either zero or two $\sigma_x$ error occur, the final state will again correspond to the noiseless phase encoding $\ket{\psi^{(2)}_{\varphi}} = U_\varphi^2 \ket{\psi^{(2)}} = (\ket{00} + e^{2 i\varphi }\ket{11})/\sqrt{2}$ for which the QFI equals $4$.  Whereas, if only a single $\sigma_x$ occurs the 
state will contain no information about the phase at all. As a result the final QFI reads $F^{(2)}_{\t{E}} = 4 (p^2 + (1-p)^2)  \geq F^{(2)}_{\t{C}}$.
Interestingly, this result may be further improved via a simple adaptive strategy.
The protocol involves entangling the initial single probe qubit with a single ancillary qubit, so that the input state is again $\ket{\psi^{(2)}}$. After a single action of the channel, $\Lambda_\varphi \otimes \mathcal{I}$,
an error correction operation is performed, where we check if a $\sigma_x$ error occurred and correct the error accordingly.
Then the channel acts  on the probe state again, and with probability $p$
yields the ideal state $\ket{\psi^{(2)}_{\varphi}}$, while if another $\sigma_x$ error occurs, the final unitary rotation $U_\varphi$ removes all the phase information from the state. Consequently, the protocol yields a QFI equal to $4p$. This protocol is actually the optimal one provided $p \geq 0.5$. If $p < 0.5$, then one simply needs to modify the recovery operation in a way that instead of correcting a single $\sigma_x$ error on the probe system the $\sigma_x$ operation is applied to the ancillary qubit. In the end the optimal QFI reads
$F^{(2)}_{\t{AD}} = 2(1 + |1-2p|) \geq F^{(2)}_{\t{E}}$ (see Appendix~\ref{app:example} for details). 

With this example in mind, one may wonder how to prove that the actual protocols are indeed optimal and what is (if any) the potential benefit of using even more general causal superposition strategies, $F^{(2)}_{\t{CS}}=?$. For larger $n$ the task becomes even more challenging, and no brute-force optimization approach can 
tell what happens in the asymptotic limit $n \rightarrow \infty$. The methods developed in this paper allow to answer all these questions.

\paragraph*{State-of-the-art bounds.}
The most powerful state-of-the-art bounds for the performance of (E) as well as (AD) strategies,
are based on the concept of minimization of certain operator norm expressions over different Kraus representation of the channel $\Lambda_\varphi = \sum_k K_{\varphi, k} \cdot K_{\varphi, k}^\dagger $ \cite{Fujiwara2008,Escher2011,Demkowicz-Dobrzanski2012,Kolodynski2013,Demkowicz-Dobrzanski2014,Sekatski2016,Demkowicz-Dobrzanski2017,Zhou2017,Zhou2020}---in what follows we drop subscript $\varphi$ in Kraus operators for conciseness.
For (E) strategies, the upper bound on the achievable QFI, reads:

\begin{equation}
\label{eq:parbound}
F^{(n)}_{\t{E}} \leq \min\limits_{\{K_k\}} 4 \left[ n \| \alpha \| + n(n-1) \| \beta\|^2   \right],
\end{equation}
where $\|\cdot \|$ denotes the operator norm, $\alpha =\sum_k \dot{K}_{k}^\dagger \dot{K}_{k}$,
$\beta = \sum_k \dot{K}_{k}^\dagger K_{k}$ and $\dot{K_{k}} = \partial_{\varphi} K_{k}$.
If a Kraus representation exists for which $\beta = 0$, the QFI scales asymptotically at most linearly with $n$---SS models---and the optimal quantum enhancement amounts to a constant factor improvement \cite{Escher2011, Demkowicz-Dobrzanski2012, Kolodynski2013, Demkowicz-Dobrzanski2014}.
If no such representation exists, then the HS 
can be preserved asymptotically \cite{Zhou2017, Zhou2020}.
Interestingly, the above bound has been proven to be asymptotically tight for both SS ($\beta = 0$) and HS ($\beta \neq 0$) models \cite{Zhou2020}.

Moving to (AD) strategies, the best state-of-the-art universally valid bound reads \cite{Demkowicz-Dobrzanski2014, Sekatski2016, Zhou2020}:
\begin{equation}
\label{eq:adbound}
F^{(n)}_{\t{AD}} \leq \min\limits_{\{K_i\}} 4 \left[ n \| \alpha \| + n(n-1) \|\beta\| \left( \|\beta \| + 2 \sqrt{\|\alpha \|}\right)\right].
\end{equation}
It is asymptotically equivalent to the parallel bound, Eq.~\eqref{eq:parbound}, in case of SS models ($\beta=0$), and, since the parallel bound is asymptotically saturable, this implies no asymptotic advantage of (AD) strategies over (E).
Still, the bound leaves space for improvement for finite $n$ and does not exclude an asymptotic advantage for HS models---the term quadratic in $n$ has a larger coefficient than the one in Eq.~\eqref{eq:parbound}.

\emph{Iterative bound.}
Below, we derive a tighter adaptive bound than the one given above, and prove it is asymptotically equivalent to the parallel one---consequently, this implies no asymptotic advantage of (AD) over (E) for all models (both SS and HS).

Let ${\Lambda}_\varphi^{(n)}(\cdot) = \sum_{\boldsymbol{k^{(n)}}}{K}_{\boldsymbol{k^{(n)}}} \cdot {K}^{\dagger }_{\boldsymbol{k^{(n)}}}$ represent a combined action of $n$ channels $\Lambda_\varphi$ in a general adaptive strategy where they are intertwined with control operations $V_i$ acting on probe and ancillary systems, as in Fig.~\ref{fig:schemes}(AD).
${K}_{\boldsymbol{k^{(n)}}}$ denote the corresponding Kraus operators, which
can be computed via the following iteration relation:
$K_{\boldsymbol{k^{(1)}}}= V_1 (K_{ k_1} \otimes \mathbb{1} )$,
\begin{equation}
\label{eq:krausiter}
K_{\boldsymbol{k^{(i+1)}}} =   V_{i+1} (K_{ k_{i+1}} \otimes \mathbb{1} ) K_{\boldsymbol{k^{(i)}}},
\end{equation} where  $\boldsymbol{k^{(i)}} = (k_i,\dots,k_1)$, 
 and $\mathbb{1}$ is acting on the ancillary system (we will drop it in what follows for conciseness of notation). 

The starting point for the 
derivation of the state-of-the art bounds as reported in Eqs.~(\ref{eq:parbound},\ref{eq:adbound}), is an observation that, given
a channel $\Lambda_{\varphi}^{(n)}$, maximization of the QFI of the output state over all inputs and sets of control operations can be
upper bounded by \cite{Fujiwara2008}:
\begin{equation}
\label{eq:generalbound}
F_{\t{AD}}^{(n)} = \max_{\rho_0, \{V_i\}} F \left[\Lambda_{\varphi}^{(n)}(\rho_0) \right] \leq \max_{\{V_i\}} \min_{\{K_{\boldsymbol{k^{(n)}}}\}}
4 \left\| \alpha^{(n)} \right\|,
\end{equation}
where $\alpha^{(n)}=\sum_{\boldsymbol{k^{(n)}}}\dot{K}^{\dagger}_{\boldsymbol{k^{(n)}}}\dot{K}_{\boldsymbol{k^{(n)}}}$, the minimization is performed over all equivalent Kraus representations of  $\Lambda_\varphi^{(n)}$ and
$\| \cdot \|$ is the operator norm. Note that for large enough ancillary system
the inequality becomes equality.
As such, this inequality is not of much practical use due to infeasibility of performing the minimization over all Kraus representations for larger values of $n$, as well as the need to  additionally perform the optimization over the control operations $V_i$. The usefulness of this inequality stems from the fact, that it is possible to further upper bound the r.h.s. of Eq.~\eqref{eq:generalbound} with norms of operators defined in terms of Kraus operators of the \emph{elementary channel}
$\Lambda_\varphi$. This is how bounds (\ref{eq:parbound},\ref{eq:adbound}) were obtained \cite{Escher2011, Demkowicz-Dobrzanski2012, Demkowicz-Dobrzanski2014, Sekatski2016, Zhou2020}.

In what follows we provide a novel step-by-step approach, where at each step we bound the maximal \emph{increase} in the final QFI thanks to the additional usage of a single quantum channel \footnote{similar philosophy may be found in \cite{Pereira2021, Wan2022}, yet the final results obtained there lacked either generality or tightness}. 
Using Eq.~\eqref{eq:krausiter} we have
\begin{multline}
\alpha^{(i+1)} = \sum_{k_{i+1},\boldsymbol{k^{(i)}}} \left({K}^{\dagger}_{\boldsymbol{k^{(i)}}}\dot{K}_{k_{i+1}}^{ \dagger}  +
 \dot{K}^{\dagger}_{\boldsymbol{k^{(i)}}} {K}_{k_{i+1}}^{\dagger} \right) \times \textrm{h.c.} \\
= \sum_{\boldsymbol{k^{(i)}}} {K}^{\dagger}_{\boldsymbol{k^{(i)}}} \alpha {K}_{\boldsymbol{k^{(i)}}} +
 {K}^{\dagger}_{\boldsymbol{k^{(i)}}} \beta \dot{K}_{\boldsymbol{k^{(i)}}} +  \dot{K}^{\dagger}_{\boldsymbol{k^{(i)}}} \beta^{\dagger} {K}_{\boldsymbol{k^{(i)}}}+
 \alpha^{(i)}.
\end{multline}
We will now use the following operator norm inequality (see Appendix~\ref{app:opnorm} for the proof):
\begin{equation}
\label{eq:opnormineq}
\left\|\sum_k L_k^\dagger A Q_k \right\|  \leq \sqrt{\left\|\sum_{k} L_k^\dagger L_k \right\|} \|A\| {\sqrt{\left\|\sum_{k} Q_k^\dagger Q_k \right\|}},
\end{equation}
which, together with the triangle inequality and the trace preservation condition, $\sum_{\boldsymbol{k^{(i)}}}  {K}^{\dagger}_{\boldsymbol{k^{(i)}}}{K}_{\boldsymbol{k^{(i)}}}=\openone$, yields:
\begin{equation}
\label{eq:iter}
\| \alpha^{(i+1)}\| \leq \|\alpha^{(i)}\| + \| \alpha\| + 2 \| \beta \| \sqrt{\|\alpha^{(i)}\|}.
\end{equation}

Let us define the following iteration,  
\begin{equation}
\label{eq:itermin}
a^{(i+1)} = a^{(i)} + \| \alpha\| + 2 \| \beta \|  \sqrt{a^{(i)}} , \quad a^{(0)}=0,
\end{equation}
which, in light of Eq.~(\ref{eq:generalbound},\ref{eq:iter}), yields $F_{\t{AD}}^{(n)}\leq 4a^{(n)}$. The resulting bound $4a^{(n)}$ may be optimized over the choice of Kraus representation of the elementary channel in each iteration \emph{separately} (how to efficiently implement this iteration numerically is described in Appendix~\ref{app:iter}) or, in a weaker variant, over a \emph{single} Kraus representation identically used in each step (for which the resulting bound will also be valid 
for (CS) strategies---see Appendix~\ref{app:causal} for the proof). Since $a^{(n)}$ is strategy-independent, the maximization over $\{V_i\}$, or, more generall, over all (CS) strategies, is no longer necessary. This finally yields
\begin{equation}
\label{eq:ADCSbounds}
F^{(n)}_{\t{AD}} \leq \min_{\{K_k\}^{\times n}}4 a^{(n)}, \quad F^{(n)}_{\t{CS}} \leq \min_{\{K_k\}}4 a^{(n)}.
\end{equation}
Interestingly, the possibility to use a different Kraus representation for each channel use allows to tighten the bound also for parallel strategies, see Appendix~\ref{app:parallel}.

\emph{Closed formula bounds.}
In order to appreciate how much tighter the obtained bounds are compared to the state-of-the-art bounds, we will provide some closed formulas for the bounds that result from a relaxed variants of the iteration procedure.
First, observe that from Eq.~\eqref{eq:opnormineq} we get $\| \beta \| \leq \sqrt{\| \alpha\|}$. From Eq.~\eqref{eq:itermin} it then follows that $a^{(n)} \leq n^2 \|\alpha\|$ (the bound obtained in \cite{Pereira2021}), which when put back into the iteration formula results in 
\begin{equation}
F^{(n)}_{\t{AD},\t{CS}} \leq \min_{\{K_k\}} 4\left(  n \| \alpha\| + n(n-1) \|\beta\| \sqrt{\|\alpha\|}  \right).
\end{equation}
Note, that the bound is noticeably tighter than Eq.~\eqref{eq:adbound} and is also valid for (CS) strategies, as the same Kraus representation is used in each step.
We also see that the difference between this bound and Eq.~\eqref{eq:parbound}
amounts to replacing one $\| \beta \|$ with $\sqrt{\|\alpha\|}$. It might be tempting to conjecture that this
difference represents is fact the asymptotic gain of (AD) over (E) strategies. This is not the case, however, as we demonstrate below.

For any fixed $\|\alpha\|, \|\beta\|$ consider the following function $f(n)= n \|\alpha\| +  n (n-1)\|\beta\|^2 + n \log n(\|\alpha\|-\|\beta\|^2)$. For $n\geq 0$ 
it can be shown (see Appendix~\ref{app:asymp}) that
$f(n+1) \geq f(n) + \|\alpha\| + 2\|\beta\| \sqrt{f(n)}$. 
Hence, in light of Eq.~\eqref{eq:itermin} we get $f(n)\geq a^{(n)}$ and as a result:
\begin{equation}
\label{eq:assymptoticbound}
F^{(n)}_{\t{AD},\t{CS}} \leq \min_{\{K_k\}}4 \left[ n \|\alpha\| + n (n-1) \| \beta\|^2 \left(1 + \frac{c \log n}{n-1} \right)\right],
\end{equation}
where $c = (\|\alpha\| - \|\beta\|^2)/\|\beta\|^2$. Since we know that the parallel bound, Eq.~\eqref{eq:parbound}, is asymptotically saturable this implies that:
\begin{equation}
\lim_{n \rightarrow \infty}\left({F^{(n)}_{\t{AD},\t{CS}}}/{F^{(n)}_{\t{E}}}\right) = 1
\end{equation}
and, hence, there is \emph{no asymptotic advantage} of (AD) nor (CS) over (E).

Interestingly, lack of asymptotic advantage thanks to adaptiveness has also been demonstrated for continuous-time models \cite{Wan2022}, a result which can 
be regarded as a limiting case of the theory we develop here (see Appendix~\ref{app:continuous} for details).

\emph{Examples.}
\begin{figure}[t!]
\includegraphics[width=0.95\columnwidth]{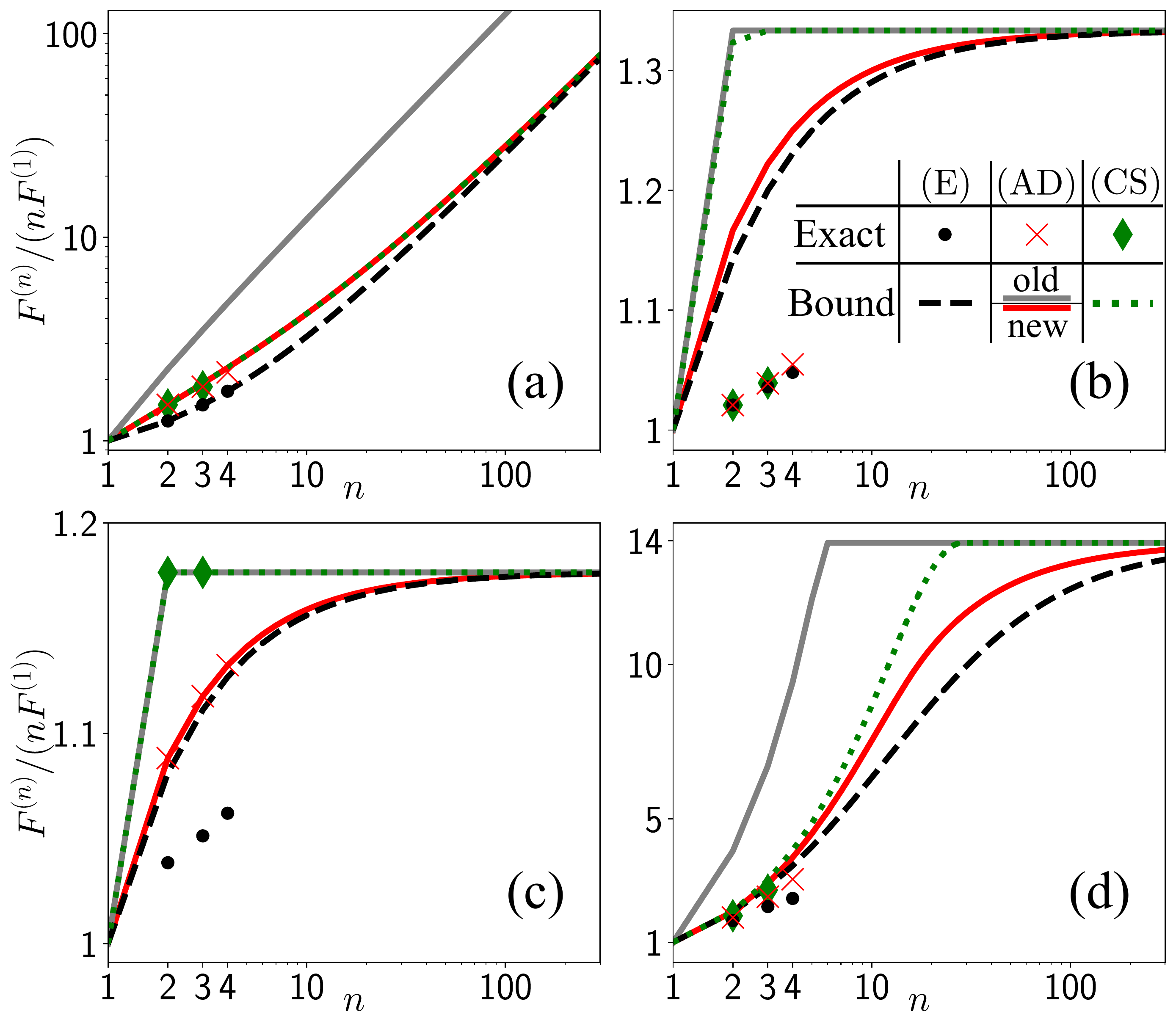}
\caption{Achievable QFI as a function of number of channels probed for parallel (E, black) , adaptive (AD, red), and causal superposition strategies (CS, green) normalized by 
$n$ times the single-channel QFI. Points represent the result of the exact optimization, while curves represent the respective bounds. 
The best previously known adaptive bound (gray) is depicted for comparison.
The four plots correspond to different metrological models with a qualitatively different behaviour: (a) dephasing perpendicular to the signal, Eq.~\eqref{eq:krausdephaseperp} ($p=0.75$); (b) dephasing parallel to the signal, Eq.~\eqref{eq:krausdephasepar} ($p=0.75$); (c) damping perpendicular to the signal, Eq.~\eqref{eq:krausdampingperp} ($p=0.15$); (d) damping parallel to the signal, Eq.~\eqref{eq:krausdampingpar} ($p=0.75$).}
\label{fig:plots}
\end{figure}
In order to illustrate the practical applications of the bounds, we compute them for four representative models and compare the results with the actual performance of the optimal protocols that can be determined numerically for small number of channel uses ($n \leq 4$) via semidefinite programming (SDP) as described in \cite{Demkowicz-Dobrzanski2014} (parallel strategies), \cite{Altherr2021} (adaptive protocols) and \cite{Liu2022f} (causal superposition protocols).
The results are presented in Fig.~\ref{fig:plots}. As a figure-of-merit we plot the achievable QFI with $n$ uses of a channel 
normalized by $n$ times $F^{(1)}$ (the maximal QFI for single-channel sensing with a possible use of ancillary systems).

Fig.~\ref{fig:plots}(a) presents results corresponding to the introductory example of perpendicular dephasing model,  Eq.~\eqref{eq:krausdephaseperp}---in all the models that follow we also assume the convention that $K_{\varphi,k} = U_\varphi K_k$ (signal comes after noise). Among the four models presented, this is the only one that admits asymptotic HS---hence the linear increase of the figure of merit. Interestingly, the bounds are saturated for $n=2$ and the optimal QFI values are equal to the ones obtained for the protocols discussed in the introductory example, proving they are indeed optimal. For larger $n$, the bounds are very tight, and, as expected, the bounds for (AD) and (CS) converge asymptotically to the (E) bound (unlike the state-of-the-art bound).

Results depicted in Fig.~\ref{fig:plots}(b) refer to the parallel dephasing model (both the unitary encoding and the dephasing are with respect to the $z$ axis), where the Kraus operators read:
\begin{equation}
\label{eq:krausdephasepar}
K_{1} = \sqrt{p} \openone, \quad K_{\varphi, 2} = \sqrt{1-p} \sigma_z.
\end{equation}
In this case, gains due to adaptiveness or causal superpositions are very modest, and the bounds are not particularly tight for low $n$---still, thanks to the general theorem, we know they are tight asymptotically.

Fig~\ref{fig:plots}(c) illustrates results for the perpendicular amplitude damping model (unitary encoding with respect to the $z$ axis, 
amplitude damping with respect to the $x$ axis): 
\begin{equation}
\label{eq:krausdampingperp}
K_{1} = \ket{-}\bra{-} + \sqrt{p} \ket{+}\bra{+}, \quad K_{2} = \sqrt{1-p}   \ket{-}\bra{+},
\end{equation}
where $\ket{\pm}=(\ket{0} \pm \ket{1})/\sqrt{2}$ are the eigenvectors of $\sigma_x$. This model is of particular interest as the finite-$n$ bounds
are saturated here both for (AD) and (CS) for all $n$. This suggests that it is highly unlikely that any tighter metrological bounds can be derived solely from the properties of the single-channel Kraus operators.

Finally, Fig.~\ref{fig:plots}(d) depicts results for the parallel amplitude damping model with:
\begin{equation}
\label{eq:krausdampingpar}
K_{1} =\ket{0}\bra{0} + \sqrt{p} \ket{1}\bra{1}, \quad K_{2} = \sqrt{1-p}   \ket{0}\bra{1}.
\end{equation}
This model illustrates particulary well how much tighter the novel bounds are, when compared with the previous state-of-the-art ones.

\emph{Conclusions and open problems.}
With the results presented in this paper, we dare to say that the theory of single-parameter quantum metrology in presence of uncorellated noise is now complete. Universal asymptotically saturable bounds are known as well as efficiently computable bounds in the regime of finite (but potentially large) number of channel uses.
This, together with exact algorithms to find optimal protocols for small $n$, provides a complete landscape of achievable quantum enhancement in realistic quantum metrology.
This said, we need to admit that in case of multiparameter models \cite{Szczykulska2016, Albarelli2022}, Bayesian models \cite{Hall2012, Rubio2020}, and most importantly models involving temporally or spatially correlated noise~\cite{Matsuzaki2011,Chin2012,Smirne2016,Beaudoin2018,Altherr2021} the quest for a full understanding of quantum metrological potential is still not complete.
Nevertheless, these achievements compare favourably to  the ones obtained in the related field of (binary) quantum channel discrimination~\cite{Pirandola2019,Katariya2020a}.
Interestingly, adaptive strategies are not useful asymptotically for asymmetric hypothesis testing~\cite{Wilde2020,wangResourceTheoryAsymmetric2019a,Fang2020}, while an advantage is possible in the symmetric setting~\cite{Harrow2010,Salek2022}.
However, easily computable asymptotic bounds, as well as practical strategies to attain them for arbitrary channels are still missing, unlike in quantum metrology.
Moreover, the asymptotic analysis of causal superposition strategies for quantum channel discrimination~\cite{Bavaresco2021,Bavaresco2022} is still an open question.

\emph{Acknowledgements.} 
This work was supported by the National Science Center (Poland) grant No.2020/37/B/ST2/02134.
FA acknowledges financial support from MUR under the ``PON Ricerca e Innovazione 2014-2020'' project EEQU.
\bibliography{adaptive-metro-biblio-PRL}

\appendix

\section{Optimal estimation strategies in case of the perpendicular dephasing model ($n=2$).}
\label{app:example}
Before presenting the analysis of the example, let us recall the definition of the QFI for completeness of the presentation. Given a parameter dependent state $\rho_\varphi$, the corresponding QFI reads:
\begin{equation}
\label{eq:sldqfi}
    F(\rho_{\varphi}) =
    \t{Tr}\left(\rho_{\varphi} L_\varphi^2 \right), \quad \frac{d\rho_\varphi}{d \varphi} = \frac{1}{2}\left(\rho_\varphi L_\varphi + L_\varphi \rho_\varphi  \right),
\end{equation}
where $L_\varphi$ is the symmetric logarithmic derivative (SLD) and can be computed from the rightmost formula above. For pure states, the formula simplifies to 
\begin{equation}
F(\ket{\psi_\varphi}) = 4 \left(  \braket{\dot{\psi}_{\varphi}| \dot{\psi}_{\varphi}}
- |\braket{\dot{\psi}_{\varphi}| \psi_{\varphi}}|^2
\right),
\end{equation}
where $\ket{\dot{\psi}_{\varphi}} = \frac{d\ket{\psi_\varphi}}{d \varphi}$.
Operationally, the inverse of the QFI provides a lower bound on the minimal achievable variance of the estimated parameter, via the quantum Cram{\'e}r-Rao bound \cite{Helstrom1976, Braunstein1994}, $\Delta^2\tilde{\varphi} \geq 1/F$.

In particular for ideal phase encoding, where 
the phase is coherently imprinted $k$ times on a balanced superposition state: 
$\ket{\psi_\varphi} = (\ket{0} + e^{i k \varphi} \ket{1})/\sqrt{2}$, $F= k^2$.
This implies, the result $F^{(1)}=1$, for single use, $n=1$, of the channel in the example discussed in the main text.

Given two uses of a channel, $n=2$, 
and utilizing no entanglement or adaptiveness, the most straightforward strategy is to 
use the two channels independently, and send optimal single qubit probes to each of them. 
In this case we get the QFI $F^{(2)}_{\t{C}}=2$ (twice the single-channel QFI). 



This value may be improved by considering a parallel strategy involving an entangled input state $\ket{\psi^{(2)}}= (\ket{00} + \ket{11})/\sqrt{2}$ (the N00N state), which results in the output state of the form:
\begin{multline}
\rho^{(2)}_{\varphi, \t{E}} = \Lambda_\varphi^{\otimes 2}\left(\ket{\psi^{(2)}}\bra{\psi^{(2)}} \right)= \\
[p^2 + (1-p)^2] \ket{\psi^{(2)}_\varphi}\bra{\psi^{(2)}_\varphi} + 2p(1-p) \ket{\phi^{(2)}}\bra{\phi^{(2)}},
\end{multline}
where $\ket{\psi^{(2)}_{\varphi}} = (\ket{00} + e^{2 i\varphi }\ket{11})/\sqrt{2}$,
is the state corresponding to an effectively noiseless phase encoding (when two or none $\sigma_x$ errors occur),
while the orthogonal state $\ket{\phi^{(2)}} = (\ket{01} + \ket{10})/\sqrt{2}$, contains no $\varphi$ information (when one $\sigma_x$ error occurred).
As a result the QFI reads $F^{(2)}_{\t{E}} = (p^2 + (1-p)^2) \times 4 \geq 2$---larger than when the two channels are used independently. 

Interestingly, this result may be further improved via a simple adaptive strategy.
The protocol involves entangling the initial single probe qubit with a single ancillary qubit in the same state as in the parallel strategy: $\ket{\psi^{(2)}}$. After a single action of the channel, $\lambda_\varphi \otimes \mathcal{I}$,
an error correction operation is performed $\mathcal{R}$, where we check if a $\sigma_x$ error occurred by projecting the state on one of the two subspaces $\mathcal{C} = \mathrm{span} \left( \ket{00},\ket{11} \right) $ and $\mathcal{E} = \mathrm{span} \left( \ket{10},\ket{01} \right)$, and correct the error accordingly.
Afterwards, the second action of the channel is applied. The resulting state at the end of the protocol reads:
\begin{multline}
\label{eq:adoutput}
\rho^{(2)}_{\varphi, \t{AD}} = (\Lambda_\varphi \otimes \mathcal{I}) \circ \mathcal{R} \circ (\Lambda_\varphi \otimes \mathcal{I}) \left(\ket{\psi^{(2)}}\bra{\psi^{(2)}} \right)= \\
p \ket{\psi^{(2)}_\varphi} \bra{\psi^{(2)}_\varphi} + (1-p)\ket{\phi^{(2)}}\bra{\phi^{(2)}}.
\end{multline}
This protocols yield $F^{(2)}_{\t{AD}} = 4p$ and is optimal provided $p \geq 0.5$. If $p < 0.5$, then one simply needs to modify the recovery
operation in a way that instead of correcting a single $\sigma_x$ error on the probe system the $\sigma_x$ operation is applied to the ancillary qubit. This yields effectively $F^{(2)}_{\t{AD}} = 4 (1-p)$. So in the end the optimal QFI reads
$F^{(2)}_{\t{AD}} = 2(1 + |1-2p|)$, which is always larger than the parallel $F^{(2)}_{\t{E}}$ except for the $p=0.5$ case when they are equal.

Having said that, it needs to be noticed that if we restrict ourselves to utilizing coherence only, we should in principle also allow a scenario where a single probe goes through the two channels sequentially. Consider $\ket{+}$ as an input (which is the optimal choice).
As a result of going through the the two channels the output state reads:
\begin{equation}
\label{eq:seqoutput}
\rho^{(2)}_{\varphi, \t{C}} = p \ket{\psi_{2\varphi}} \bra{\psi_{2\varphi}} + (1-p)\ket{+}\bra{+}.
\end{equation}
This state looks very similar to the output state resulting form the adaptive strategy, Eq.~\eqref{eq:adoutput}. Still, in  
Eq.~\eqref{eq:adoutput} we may perform a measurement that unambiguously discriminates between the ideally phase-encoded state, $\ket{\psi_\varphi^{(2)}}$, and the non-informative state $\ket{\phi^{(2)}}$. 
In case of Eq.~\eqref{eq:seqoutput} this is no longer possible as the ``signal'' state $\ket{\psi_{2\varphi}}$
and non-informative $\ket{+}$ are not orthogonal in general and hence one cannot simply filter out the non-informative term. Indeed, direct computation of the QFI, using Eq.~\eqref{eq:sldqfi}, for the state $\rho_{\varphi,C}^{(2)}$ yields
\begin{equation}
F^{(2)}_{\t{C}} = 2p[1+p+(1-p)\cos2 \varphi].
\end{equation}
While in general this value is smaller than $F^{(2)}_{\t{AD}}$, it approaches the performance of the adaptive scheme for $\varphi \approx 0$, and $p \geq 0.5$, $F^{(2)}_{\t{C}} \approx 4 p$. This shows that sometimes even a basic sequential scheme  (utilizing only coherence, and no adaptiveness) may be superior to the optimal entanglement-based scheme.

It should be pointed out, however, that 
for larger $n$ the simple sequential strategy will quickly become ineffective due to a build-up of unchecked decoherence effects which cause the resulting QFI to be damped. In this regime, more advanced strategies are required (E), (AD) or (CS) in order to exploit the full potential of quantum enhanced sensing.

\section{Operator norm inequality}
\label{app:opnorm}
Here we prove Eq.~\eqref{eq:opnormineq},
\begin{equation}
\label{eq:ineqnorm}
\left\|\sum_{k=1}^K L_k^\dagger A Q_k \right\|  \leq \sqrt{\left\|\sum_{k=1}^K L_k^\dagger L_k \right\|} \|A\| {\sqrt{\left\|\sum_{k=1}^K Q_k^\dagger Q_k \right\|}},
\end{equation}
which is a key step in the derivation of the bounds. 
Let us introduce the following $K \times K$ block matrices:
\begin{equation}
\tilde{L} = \left[\begin{array}{cccc}
L_1 & 0 & \dots & 0\\
\vdots & \vdots &  &\vdots\\
L_K & 0 & \dots & 0
\end{array}\right], \ \tilde{Q} = \left[\begin{array}{cccc}
Q_1 & 0 & \dots & 0\\
\vdots & \vdots &   &\vdots\\
Q_K & 0 & \dots & 0
\end{array}\right], 
\end{equation}
and $\tilde{A} = \openone_{K \times K} \otimes A$.
Note that 
\begin{equation}
\tilde{L}^\dagger \tilde{A} \tilde{Q} =  \left[\begin{array}{cccc}
\sum_{k=1}^K L_k^\dagger A Q_k & 0 & \dots & 0\\
 \vdots &  \vdots &  & \vdots\\
0 & 0 & \dots & 0
\end{array}\right].    
\end{equation}
This implies that:
\begin{multline}
\left\|\sum_{k=1}^K L_k^\dagger A Q_k \right\| = 
\left\|\tilde{L}^\dagger \tilde{A} \tilde{Q} \right\| \leq \| \tilde{L}^\dagger\| \| \tilde{A}\| \|\tilde{Q}\|,
\end{multline}
where the inequality follows from the sub-multiplicavity property of the operator norm.
Invoking now another operator norm property valid for arbitrary operators acting on a Hilbert space,  
 $\| A\|  = \sqrt{\|A^\dagger A\|}$ and $\|A^\dagger\|=\|A\|$, we get 
\begin{equation}
 \| \tilde{L}^\dagger\| \| \tilde{A}\| \|\tilde{Q}\| = \sqrt{\| \tilde{L}^\dagger \tilde{L}\|} \| \tilde{A} \| \sqrt{\|\tilde{Q}^\dagger \tilde{Q}\|}.
\end{equation} 
Finally, noticing that $\|\tilde{L}^\dagger \tilde{L}\|= \|\sum_k L_k^\dagger L_k\|$, 
$\| \tilde{A}\| = \| A \|$ and 
$\|\tilde{Q}^\dagger \tilde{Q}\|= \|\sum_k Q_k^\dagger Q_k\|$ we
arrive at Eq.~\eqref{eq:ineqnorm}.

\section{Derivation of the bound for causal superposition strategies}
\label{app:causal}
For (CS) strategies, any quantum superposition of $n!$ possible causal orders of $n$ elementary channels can be created. This can be achieved by entangling all different causal orders with an external, $n!$-level quantum control system \cite{Araujo2014,Liu2022f}. 
As in (AD) strategies, we are allowed to put control unitaries between channels, moreover, for different causal orders, control unitaries may be different. Therefore, (AD) is a subclass of (CS), in which only one causal order of channels is probed. 
The channel describing the action of $n$ elementary channels $\Lambda$ in the superposition of different causal orders is  $\Lambda_{\t{CS}}^{(n)}(\cdot) = \sum_{\boldsymbol{k^{(n)}}} {K}^{\t{CS}}_{\boldsymbol{k^{(n)}}} \cdot {K}^{\t{CS}\dagger}_{\boldsymbol{k^{(n)}}}$, where
\begin{equation}
K_{\boldsymbol{k^{(n)}}}^{\t{CS}} = \sum_{\pi \in \sigma(n)} K^{\pi}_{\boldsymbol{k^{(n)}_\pi}} \otimes \ket{\pi}\bra{\pi},
\end{equation}
 $\sigma(n)$ is the set of all $n$-element permutations, $\boldsymbol{k^{(n)}_\pi} = (k_{\pi(n)},...,k_{\pi(1)})$.
 The Kraus operators $K^{\pi}_{\boldsymbol{k^{(n)}}}$ are defined iteratively, as in the adaptive case: $K^{\pi}_{\boldsymbol{k^{(1)}}}= V^\pi_1 (K_{ k_{1}} \otimes \mathbb{1} )$,
\begin{equation}
\label{eq:C2}
K^{\pi}_{\boldsymbol{k^{(i+1)}}} =   V_{i+1}^\pi (K_{ k_{i+1} } \otimes \mathbb{1} ) K^{\pi}_{\boldsymbol{k^{(i)}}},
\end{equation} 
$V_{i}^\pi$ is the control unitary applied after $i$-th channel when the order of channels is described by permutation $\pi$, $\ket{\pi}$ is a corresponding state of a control system. 

The (CS) family of strategies has been studied in the context of quantum metrology in Ref.~\cite{Liu2022f}, where numerical values of the QFI were obtained for $n=2$ and $n=3$ (the Authors use abbreviation ``sup'' instead of ``CS''). Notice, that the formal definition of (CS) strategies from Appendix C.4 of Ref.~\cite{Liu2022f} does not contain the explicit form of the channel $\Lambda_{\t{CS}}^{(n)}$.
The explicit Kraus representation of a similar channel is present, for example, in \cite{Chapeau-Blondeau2021} (for $n=2$).
The Choi-Jamio{\l}kowski matrix of a channel involving superpositions of different causal orders for $n>2$ is explicitly written down in \cite{Wechs2021}. 

Before we proceed further, let us show that we defined $\Lambda^{(n)}_{\t{(CS)}}$ in a way that does not depend on the choice Kraus represenations of elementary channels $\Lambda$ in \eqref{eq:C2}.
Starting from an arbitrary Kraus representation $\{K_i\}$, we can construct any equivalent representation of the same channel $\{K'_j\}$ as $K'_j = \sum_i u_{ji} K_i$, where coefficients $u_{ij}$ form a unitary matrix \cite{Fujiwara2008}. For a fixed $\pi$, the channel $\Lambda_\pi^{(n)}$, given by Kraus operators $K^\pi_{\boldsymbol{k^{(n)}}}$, is the concatenation of elementary channels $\Lambda$ and unitary operations $V^\pi_1$, $V^\pi_2$,...,$V^\pi_n$--- $\Lambda_\pi^{(n)}$ is well defined without referring to a particular  Kraus representations of the elementary channel $\Lambda$. For this reason, changing the representation of each $\Lambda$ may only change the Kraus operators $\{K^\pi_{\boldsymbol{k^{(n)}}}\}$ to an equivalent representation $\{K_{\boldsymbol{l^{(n)}}}^{'\pi}\}$, where $ K_{\boldsymbol{l^{(n)}}}^{'\pi} = \sum_{\boldsymbol{k^{(n)}}} u^{\pi}_{\boldsymbol{l^{(n)}}, \boldsymbol{k^{(n)}}} K^\pi_{\boldsymbol{k^{(n)}}}$, the coefficients  $u^{\pi}_{\boldsymbol{l^{(n)}}, \boldsymbol{k^{(n)}}}$ form a unitary matrix. Consequently, the Kraus operators $\{K_{\boldsymbol{k^{(n)}}}^{\t{CS}}\}$ transform to $\{K_{\boldsymbol{l^{(n)}}}^{'\t{CS}}\}$, where
\begin{equation}
K_{\boldsymbol{l^{(n)}}}^{'\t{CS}} = \sum_{\pi \in \sigma(n)} \left(  \sum_{\boldsymbol{k^{(n)}}} u^{\pi}_{\boldsymbol{l^{(n)}}, \boldsymbol{k^{(n)}}} K^\pi_{\boldsymbol{k^{(n)}}}\right) \otimes \ket{\pi}\bra{\pi}.
\end{equation}
This can be written as
\begin{equation}
K_{\boldsymbol{l^{(n)}}}^{'\t{CS}} = \sum_{\boldsymbol{k^{(n)}}} u_{\boldsymbol{l^{(n)}}, \boldsymbol{k^{(n)}}} K^{\t{CS}}_{\boldsymbol{k^{(n)}}},
\end{equation}
where 
\begin{equation}
u_{\boldsymbol{l^{(n)}}, \boldsymbol{k^{(n)}}} = \sum_\pi u^\pi_{\boldsymbol{l^{(n)}}, \boldsymbol{k^{(n)}}} \otimes \ket{\pi} \bra{\pi}
\end{equation}
is a unitary matrix. Therefore, the Kraus operators $K_{\boldsymbol{l^{(n)}}}^{'\t{CS}}$ form an equivalent representation of a channel $\Lambda^{(n)}_\t{CS}$, so physical properties of this channel do not depend on the choice of represantions of $\Lambda$.

In what follows, we derive an upper bound for the QFI achievable with any (CS) strategy involving $n$ elementary channels, which can be efficiently computed for large $n$. We know that 
\begin{equation}F^{(n)}_{\t{CS}} \le \underset{\left\{K^\t{CS}_{\boldsymbol{k^{(n)}}}\right\}}{\t{min}} \left\| \alpha^{(n)}_{\t{CS}} \right\|,\quad \alpha^{(n)}_{\t{CS}} = \sum_{\boldsymbol{k^{(n)}}} \dot K^{\t{CS} \dagger}_{\boldsymbol{k^{(n)}}} \dot K^{\t{CS}}_{\boldsymbol{k^{(n)}}},
\end{equation}
where the minimization is taken over all Kraus representations of the channel $\Lambda_{\t{CS}}^{(n)}$.
Such minimization is very hard to perform, but we can obtain a less tight but more feasible upper bound by minimizing over representations of a single channel only, assuming that the representation of each elementary channel $\Lambda$ is the same:
\begin{equation}
\label{C4}
F^{(n)}_{\t{CS}} \le \underset{\{K_k\}}{\t{min}} \left\| \alpha^{(n)}_{\t{CS}} \right\|.
\end{equation}
For this family of representations, we have
\begin{equation}
\alpha_\text{CS}^{(n)} = \sum_{\boldsymbol{k^{(n)}}} \sum_{\pi \in \sigma(n)} \dot K^{\pi \dagger}_{\boldsymbol{k^{(n)}_\pi}} \dot K^{\pi}_{\boldsymbol{k^{(n)}_\pi}} \otimes \ket{\pi} \bra{\pi}.
\end{equation}
The matrix $\alpha_\text{CS}^{(n)}$ has a block diagonal structure, where different blocks correspond to different $\pi$ because the states $\ket{\pi}$ are orthogonal to each other. Therefore,
\begin{equation}
\label{C6}
\left\| \alpha_\text{CS}^{(n)} \right\| = \underset{\pi \in \sigma(n)}{\text{max}} \left\| \sum_{\boldsymbol{k^{(n)}}} \dot K^{\pi \dagger}_{\boldsymbol{k^{(n)}_\pi}} \dot K^{\pi}_{\boldsymbol{k^{(n)}_\pi}} \right\|.
\end{equation}
The summation inside a norm runs over all possible vectors $\boldsymbol{k^{(n)}}$, so it does not change when lower indices $\boldsymbol{k_\pi^{(n)}}$ are replaced with $\boldsymbol{k^{(n)}}$, so we have
\begin{equation}
\label{C7}
\left\| \alpha_\text{CS}^{(n)} \right\| = \underset{\pi \in \sigma(n)}{\text{max}} \left\| \sum_{\boldsymbol{k^{(n)}}} \dot K^{\pi \dagger}_{\boldsymbol{k^{(n)}}} \dot K^{\pi}_{\boldsymbol{k^{(n)}}} \right\|
\end{equation}
In this step, the assumption that the Kraus representations of all $n$ elementary channels are the same, was crucial---otherwise, a different $\pi$ would correspond to a different order of representations, so \eqref{C7} would not follow from \eqref{C6}. That is why we cannot minimize over different representations for different channels, as we did for (AD) strategies, and the (CS) bound is less tight.

For a fixed $\pi$, the Kraus operators $K^{\pi}_{\boldsymbol{k^{(n)}}}$ describe an adaptive strategy on $n$ elementary channels with control unitaries $V_{i}^\pi$. In \eqref{C7}, maximization over $\pi$ is equivalent to maximization over different sets of control unitaries $V_{i}^\pi$ with a fixed causal order of channels, so we have
\begin{equation}
\label{C9}
\left\| \alpha_\text{CS}^{(n)} \right\| \le \underset{\{V_i\}}{\text{max}} \left\| \sum_{\boldsymbol{k^{(n)}}} \dot K^{ \dagger}_{\boldsymbol{k^{(n)}}} \dot K_{\boldsymbol{k^{(n)}}} \right\|,
\end{equation}
where the maximization is over all sets of control unitaries, and $K_{\boldsymbol{k^{(n)}}}$ for a given $\{V_i\}$ are defined in \eqref{eq:krausiter}. The upper bound for the r.h.s.  for a fixed elementary channel representation $\{K_k\}$ is
\begin{equation}
\label{C10}
 \underset{\{V_i\}}{\text{max}} \left\| \sum_{\boldsymbol{k^{(n)}}} \dot K^{ \dagger}_{\boldsymbol{k^{(n)}}} \dot K_{\boldsymbol{k^{(n)}}} \right\| \le a^{(n)},
\end{equation}
which follows directly from the derivation of the (AD) bound.  Finally, we can minimize both sides of \eqref{C9} and \eqref{C10} over $\{K_k\}$ to obtain
\begin{equation}
\underset{\{K_k\}}{\t{min}} \left\| \alpha_\text{CS}^{(n)} \right\| \le \underset{\{K_k\}}{\text{min}} \underset{\{V_i\}}{\text{max}} \left\| \sum_{\boldsymbol{k^{(n)}}} \dot K^{ \dagger}_{\boldsymbol{k^{(n)}}} \dot K_{\boldsymbol{k^{(n)}}} \right\| \le \underset{\{K_k\}}{\text{min}} a^{(n)}.
\end{equation}
This inequality, together with \eqref{C4}, leads directly to the upper bound for $F^{(n)}_{\t{CS}}$ from \eqref{eq:ADCSbounds}.

 The considered class of strategies (CS) is quite general, since it allows superpositions of all different causal orders of channels combined with all possible intermediate controls---moreover, for different causal orders, unitary controls may be different. However, quantum mechanics in principle allows for even more general non-causal strategies, the most general class of strategies is schematically shown in Fig. 1.e in Ref.~\cite{Liu2022f}, where the Authors refer to it using abbreviation ``ICO''. The advantage of (ICO) over (CS) strategies is usually very small, but possible to demonstrate numerically already for $n=2$ \cite{Liu2022f}. Therefore, it is interesting to ask, whether our newly derived bound can be applied to the most general non-causal strategies (ICO). Unfortunately, the answer is negative, and a counterexample exists already for $n=3$. Let us consider a single qubit channel studied in the \textit{introductory example}, where  noise is described by the Kraus operators from \eqref{eq:krausdephaseperp} with $p=0.75$. For $n=3$, exact values of the maximal QFI associated with (CS) and (ICO) strategies (obtained using procedures from \cite{Liu2022f}) are $F_{\t{CS}}^{(3)} = 5.52$, $F_{\t{ICO}}^{(3)} = 5.84$, whereas the upper bound for QFI for (CS) strategies is $\min_{\{K_k\}}4 a^{(3)} = 5.73 $, so we have $F_{\t{CS}}^{(3)}< \min_{\{K_k\}}4 a^{(3)} < F_{\t{ICO}}^{(3)} $. Therefore, the problem of finding a feasible upper bound on the QFI for the most general non-causal strategies is still open. However, it is not clear whether all (ICO) strategies are possible to implement---probably this fundamental question should be addressed first.

\section{Computing the iterative bounds}
\label{app:iter}
Firstly, let us describe more precisely the construction of the (AD) bound. When we construct $a^{(n)}$, defined in \eqref{eq:itermin}, we can choose the Kraus representation of the elementary channel $\{K_k\}$ independently in each step, so the norms $\|\alpha\|$, $\|\beta\|$ are different in each step. Then, we want to minimize $a^{(n)}$ over all possible sets of $n$ represenations $\{K_k\}^{\times n}$, the $i$-th element of such a set corresponds to the representation chosen in $i$-th step. However,  $a^{(i+1)}$ depends only on $a^{(i)}$ and the representation chosen in the step $i+1$, moreover, $a^{(i+1)}$ is an increasing function of $a^{(i)}$.   Therefore, the minimum of $a^{(i+1)}$ over $\{K_k\}^{\times (i+1)}$ can be obtained by inserting the minimum of $a^{(i)}$ over $\{K_k\}^{\times i}$ into \eqref{eq:itermin}, and then minimizing the resulting expression over the representation of $(i+1)$-th channel. 

As a result, the full optimization over $n$ Kraus representations used in $n$ steps may be easily obtained by performing $n$ iterative optimizations over a single Kraus representation in each successive step, namely:
\begin{equation}
\label{tildea}
\begin{split}
&\min_{\{K_k\}^{\times n}} a^{(n)} = \tilde{a}^{(n)},\quad \t{where}\quad
\tilde{a}^{(0)} = 0, \\
&\tilde{a}^{(i+1)} = \underset{\left\{ K_k \right\}}{\t{min}}~ \left[\tilde{a}^{(i)} + \left\| \alpha_{\{K_k\}} \right\| + 2 \left\| \beta_{\{K_k\}}  \right\| \sqrt{\tilde{a}^{(i)}}\right].
\end{split}
\end{equation}
We used notation $\alpha_{\{K_k\}}$ and $\beta_{\{K_k\}}$ to indicate the dependence of an elementary channel representation $\{K_k\}$, over which we minimize in step $(i+1)$. 
We are now ready to demonstrate the numerical procedures for computing (AD) and (CS) bounds.
\subsection{Minimizing $\|\alpha\|$ assuming $\|\beta\| \le b$}
Let us show how to formulate the following minimization problem
\begin{align}
\label{eq:gdef}
g(b)=~~&\t{minimize}_{\{K_k\}} \quad \left\| \alpha_{\{K_k\}} \right\| \\
&\t{subject to} \quad \left\| \beta_{\{K_k\}} \right\| \le b
\end{align}
as an SDP. Firstly, as shown in \cite{Fujiwara2008,Demkowicz-Dobrzanski2012}, we can significantly limit the class of Kraus representations over which we minimize. Starting from arbitrary $\{K_k\}$, it is enough to consider representations $\{\tilde K_k\}$ satisfying
\begin{equation}
\label{eq:KrausTilde}
\tilde{K}_k = K_k,\quad \dot{\tilde{K}}_k = \dot{\tilde{K}}_k - i \sum_j h_{kj} K_j
\end{equation}
to obtain all possible values of $(\|\alpha \|, \|\beta\|)$ for a given channel $\Lambda$, the coefficients $h_{kj}$ must form a hermitian matrix $h \in \t{Herm}(\mathbb{C}^{r \times r})$. We can replace the minimization over $\{K_k\}$ with a minimization over $h$ in \eqref{eq:gdef}. Following the path from \cite{Demkowicz-Dobrzanski2012, Kolodynski2013}, let us construct the following matrices
\begin{align}
\label{matA}
A &= \left( \begin{array}{c|ccc}
        \lambda \mathbb{1}_d& \dot{\tilde{ K}}_1^\dagger  & \hdots &  \dot{\tilde{ K}}_r^\dagger \\ \hline
        \dot{\tilde{ K}}_1&  &  &  	 \\
        \vdots&  & \mathbb{1}_{d \cdot r} &    \\ 
        \dot{\tilde{ K}}_r &  &  &   \end{array}\right), \\ B &= \left( \begin{array}{c|c}
        b \mathbb{1}_d& i \sum_{k=1}^r \dot{\tilde{ K}}_k^\dagger K_k  \\ \hline
        \left(i \sum_{k=1}^r \dot{\tilde{ K}}_k^\dagger K_k\right)^\dagger& b \mathbb{1}_d.  	 
 \end{array}\right) 
 \label{matB}
\end{align}
It can be shown, using Schur's complement condition, that
\begin{equation}
A \succeq 0 \iff \lambda \ge \| \alpha_{\{\tilde K_k\}} \|  ,\quad B \succeq 0 \iff b \ge \| \beta_{\{\tilde K_k\}} \|.
\end{equation}
Therefore, the constraint $B \succeq 0$ is equivalent to upper bounding $\|\beta\|$ by $b$, and minimization of  $\lambda$ is equivalent to minimization of $\| \alpha \|$.
All elements of $A$ and $B$ are linear in $\lambda $ and $h$, so we can use constraints for positivity of $A$ and $B$ in the SDP. All this observations allow us to write down the minimization problem \eqref{eq:gdef} as an SDP described in Algorithm \ref{alg1}.
\begin{algorithm}[H]
  \caption{Minimize $\| \alpha\|$ given $\|\beta\| \le b$}
  \label{alg1}
   \begin{algorithmic}[1]
   \Require  $K_k, \dot K_k \in \mathbb{C}^{d \times d}$ for $k \in \{1,...,r\}$ \Comment \texttt{Kraus operators of elementary channel $\Lambda$ and their derivatives } 
   \Require $b \ge 0$ \Comment \texttt{Upper bound for $\|\beta\|$}
   \Ensure $g(b)=\underset{\{K_k\}}{\t{min}} \|\alpha\|$ s.t. $\|\beta\| \le b$
   \State variables: $\lambda \ge 0$, $h \in \t{Herm}(\mathbb{C}^{r \times r})$ \Comment \texttt{Variables to optimize in SDP}
   \For{$k~~ \t{in} ~~(1,...,r)$}
        \State $\dot{\tilde{ K}}_k := \dot{K}_k - i \sum_{j=1}^r h_{kj}K_j$
      \EndFor 
      \State construct matrices $A,B$ defined in \eqref{matA}, \eqref{matB}
      
 \State $\t{minimize}_{\lambda, h} \quad \lambda $ \newline
 subject to \quad $A \succeq 0 $, $B \succeq 0$ \Comment \texttt{SDP} 
 \State \textbf{Output} $\lambda$
   \end{algorithmic}
\end{algorithm}

\subsection{Computing (AD) and (CS) bounds}

Let us prove the following lemma:

\textbf{Lemma 1} Let $f(x,y)$ be an increasing function of $x$ and $y$ with non-negative values.  Then, 

\begin{equation}\min_{\{K_k\}} f\left(\|\alpha_{\{K_k\}}\|, \|\beta_{\{K_k\}}\|\right) = \min_{b \in [l,r]} f\left(g(b), b\right),\end{equation}
where $l = \min_{\{K_k\}} \|\beta_{\{K_k\}}\| $, $r = \min_{\{K_k\}} \sqrt{\|\alpha_{\{K_k\}}\|} $, $g(b)$ is defined in \eqref{eq:gdef}.

\textit{Proof}
First note that
\begin{multline}
\min_{\{K_k\}} f\left(\|\alpha_{\{K_k\}}\|, \|\beta_{\{K_k\}}\|\right) = \\
\min_{b\geq l}\,\,\min_{\{K_k\}:\|\beta_{\{K_k\}}\| \le b}\,\,f\left(\|\alpha_{\{K_k\}}\|, b\right)=
\min_{b\geq l} f\left(g(b), b\right),
\end{multline}
where we used the fact that $f(x,y)$ is increasing with $y$ (1st equality) and $x$ (2nd equality).
What remains to be proven is that the minimal value is always obtained for $b\leq r$. 
 From \eqref{eq:opnormineq}, we have $\|\beta_{\{K_k\}}\| \le \sqrt{\|\alpha_{K_k}\|}$, which means that $l \le r$. Let $\{{K'}_k\}$ be the Kraus representation minimizing $\| \alpha\|$, which means that $\sqrt{a'}=\sqrt{\| \alpha_{\{{K'}_k\}}\|} = r$, $b'=\|\beta_{\{{K'}_k\}}\| \le \sqrt{\| \alpha_{\{{K'}_k\}}\|} = r$. Obviously, $a'\le a$, so when we choose $\{K_k\}$ such that $b>r$, we have $a' \le a$ and $b'=r<b$, so $f(a',b')<f(a,b)$, which means that the minimum cannot be achieved for $b>r$.$\blacksquare$

The values of $l$ and $r$ can be found using an SDP, as descibed in \cite{Demkowicz-Dobrzanski2012, Kolodynski2013}---the result of a minimization of $\lambda$ with the constraint $A \succeq 0$ is $r^2$, and the minimum of $b$ with the constraint $B \succeq 0$ equals to $l$.

Notice, that $\tilde{a}^{(i+1)}$ defined in \eqref{tildea} is an increasing function of $\|\alpha\|$ and $\|\beta\|$, so we can use Lemma 1 to perform the minimization over $\{K_k\}$:
\begin{equation}
\tilde{a}^{(i+1)} = \min_{b \in [l,r]} \left[\tilde{a}^{(i)} + g(b) + 2b \sqrt{\tilde{a}^{(i)}}\right]
\end{equation}
In practice, we do not have an analytical form of $g(b)$, and we need to solve an SDP problem for each argument $b$ independently to obtain $g(b)$. Therefore,  minimization over $b \in [l,r]$ is approximated with minimization over $b \in \t{Linspace}(l,r,p)$, where $\t{Linspace}(l,r,p)$ is a $p$-element arithmetic sequence whose first element is $l$ and last element is $r$. Notice, that this approximation can only increase the final result---therefore, even for small sampling precision $p$, we obtain valid, but not necessary tight, upper bounds. We used the value $p=500$ to generate the data presented in this paper, further precision increase did not lead to any significant difference.

To sum up, the procedure to compute the upper bound for $F^{(n)}_{\t{AD}}$ is as follows. Firstly, we compute $l$ and $r$ using the SDP. Secondly,  $g(b)$ is calculated for $b \in \t{Linspace}(l,r,p)$. Finally, we initialize $\tilde{a}^{(0)}=0$, and compute $\tilde{a}^{(i)}$ for larger $i$ iteratively using \eqref{tildea}. Notice, that the computational cost of each iteration is very low since we computed all the required values of $g(b)$ at the beginning, and we keep them in the memory. The computation of $g(b)$ for $p$ different values of $b$ is the most time-consuming part, but even for a high precision ($p=500$), the computation time did not exceed two minutes. Once we do this computation, we can easily get an upper bound for $F^{(n)}_{\t{AD}}$ even for large values of $n$. 

The procedure to compute an upper bound for $F^{(n)}_{\t{CS}}$ is very similar. Notice, that $a^{(i)}$ defined in \eqref{eq:itermin} is an increasing function of $\|\alpha\|$ and $\|\beta\|$, so we can use Lemma 1 to find its minimum over a single $\{K_k\}$:
\begin{equation}
\begin{split}
 \min_{\{K_k\}} a^{(n)} &= \min_{b \in [l,r]} \hat{a}^{(n)}(b)\quad,\t{where}\quad \hat{a}^{(0)}(b) = 0, \\
 &\hat{a}^{(i+1)}(b) = \hat{a}^{(i)}(b) + g(b) + 2b\sqrt{\hat{a}^{(i)}(b)}
 \end{split}
\end{equation}
Subsequent values of $\hat{a}^{(i)}(b)$ are found iteratively for all $b\in \t{Linspace}(l,r,p)$, and then minimization over $b$ is performed with the help of previously calculated values of $g(b)$. The whole procedure of computing bounds for $F_{\t{AD}}^{(n)}$ and $F_{\t{CS}}^{(n)}$ is summarized in Algorithm \ref{alg2}. 
\begin{algorithm}[H]
  \caption{Iterative bounds for $F_{\t{AD}}^{(n)}$ and $F_{\t{CS}}^{(n)}$}
  \label{alg2}
   \begin{algorithmic}[1]
   \Require  $K_k, \dot K_k \in \mathbb{C}^{d \times d}$ for $k \in \{1,...,r\}$ 
   \Require $n \in \mathbb{Z}_+$ \Comment\texttt{Max number of elementary channels}
   \Require $p\in \mathbb{Z}_+$    \Comment \texttt{Precision of sampling}
   \Ensure AD, CS: lists of size $n$, $\t{AD/CS}[k]$ is an upper bound for $F_{\t{AD/CS}}^{(k)}$
   \State $l := \min_{\{K_k\}}\|\beta\|$
   \State $r := \min_{\{K_k\}}\sqrt{\|\alpha\|}$
   \State variables: $\t{a}[p], \t{b}[p], \t{t1}[p],\t{t2}[p], \t{AD}[n], \t{CS}[n] $ \Comment \texttt{Lists of real numbers of size $[\cdot]$, initialized with 0s}
   \State $\t{b} := \t{Linspace}\left(l,r,p \right)$ \Comment \texttt{ $\t{b}[1]=l$, $\t{b}[p]=r$, $\t{b}[1],\t{b}[2],...,\t{b}[p]$ is arithmetic sequence}
   \For{$j~~ \t{in} ~~(1,...,p)$}
   
   $\t{a}[j] := g(\t{b}[j])$ \Comment \texttt{ Use Alg. \ref{alg1}}
   \EndFor \newline
   \Comment \texttt{Computing AD bounds}
   \State $m:=0$
   \For{$k~~ \t{in} ~~(1,...,n)$}

   \For{$j~~ \t{in} ~~(1,...,p)$}

   \State $\t{t1}[j] := m + \t{a}[j]+2*\t{b}[j]*\sqrt{m}$
    \EndFor

   \State $m := \t{Minimum}[\t{t1}]$
   \State $\t{AD}[k] := 4*m$
   \EndFor
   \newline
   \Comment \texttt{Computing CS bounds}
   \For{$k~~ \t{in} ~~(1,...,n)$}

   \For{$j~~ \t{in} ~~(1,...,p)$}

   \State $\t{t2}[j] := \t{t2}[j] + \t{a}[j]+2*\t{b}[j]*\sqrt{\t{t2}[j]}$
    \EndFor

   \State $\t{CS}[k] := 4*\t{Minimum}[\t{t2}]$
   \EndFor
   \State \textbf{Output} AD, CS
   \end{algorithmic}
\end{algorithm}
  The old bound for (AD) strategies can be also computed numerically using Lemma 1---the minimization from \eqref{eq:adbound} simplifies to:
  \begin{equation}
  \begin{split}
\min\limits_{\{K_i\}} 4 \left[ n \| \alpha \| + n(n-1) \|\beta\| \left( \|\beta \| + 2 \sqrt{\|\alpha \|}\right)\right] =    \\ \min_{b \in [l,r]} 4 \left[ n g(b) + n(n-1)b(b+2\sqrt{g(b)} \right].
\end{split}
  \end{equation}
  Analogously, we can compute the new analytical bound from \eqref{eq:assymptoticbound}, or the bound for (E) strategies \eqref{eq:parbound}. Therefore, the function $g(b)$ fully characterizes an elementary channel $\Lambda_\varphi$ from a metrological point of view---if we know $g(b)$ (at least for some values of $b$), we can easily compute all metrological bounds.
\subsection{Tightening of the bound for parallel strategies}
\label{app:parallel}
Looking at the original derivation of the bound for parallel strategies in \cite{Fujiwara2008}, we see that it may be tightened by allowing different Kraus representations for each elementary Hilbert space in Eq. (16) of \cite{Fujiwara2008} . However, the full optimization here is a bit more complicated to perform than in the adaptive case, while the advantage over the standard parallel bound \eqref{eq:parbound} is not significant for the typical examples (what we checked numerically). Still, it is worth noting that such an optimized bound is guaranteed to be tighter than the newly derived adaptive one \eqref{eq:ADCSbounds}, which we show below.

The channel describing the parallel action of $n$ elementary channels $\Lambda$ is $\Lambda_{\t{E}}^{(n)}(\cdot) = \sum_{\boldsymbol{k^{(n)}}} {K}^{\t{E}}_{\boldsymbol{k^{(n)}}} \cdot {K}^{\t{E}\dagger}_{\boldsymbol{k^{(n)}}} $, where $K^{\t{E}}_{\boldsymbol{k^{(1)}}} = K_{k_1} \otimes \mathbb{1}$,
\begin{equation}
K^{\t{E}}_{\boldsymbol{k^{(i+1)}}} = K_{k_{i+1}} \otimes K^{\t{E}}_{\boldsymbol{k^{(i)}}},
\end{equation}
$\mathbb{1}$ is acting on an ancillary system. Using this definition, we obtain the following iteration relations for $\alpha_{\t{E}}^{(n)} = \sum_{\boldsymbol{k^{(n)}}}\dot{K}^{ \t{E} \dagger}_{\boldsymbol{k^{(n)}}}\dot{K}^{\t{E}}_{\boldsymbol{k^{(n)}}}$, $\beta_{\t{E}}^{(n)} = \sum_{\boldsymbol{k^{(n)}}}\dot{K}^{ \t{E} \dagger}_{\boldsymbol{k^{(n)}}}K^{\t{E}}_{\boldsymbol{k^{(n)}}}$ :
\begin{align}
\alpha_{\t{E}}^{(i+1)} &= \alpha \otimes \mathbb{1} + \mathbb{1} \otimes \alpha_{\t{E}}^{(i)} - 2 \beta \otimes \beta_{\t{E}}^{(i)}, \\
\beta_{\t{E}}^{(i+1)} &= \beta \otimes \mathbb{1} + \mathbb{1} \otimes \beta_{\t{E}}^{(i)},
\end{align}
$\mathbb{1}$ acting on an ancillary system was omitted.
From the triangle inequality:
\begin{align}
\| \alpha_{\t{E}}^{(i+1)} \| &\le \| \alpha_{\t{E}}^{(i)} \| + \|\alpha\| + 2 \| \beta \| \| \beta_{\t{E}}^{(i)}\| , \\
\| \beta_{\t{E}}^{(i+1)}\| &\le  \| \beta_{\t{E}}^{(i)}\| + \|\beta\|.
\end{align}
Let us define
\begin{align}
\label{eq:D18}
a_{\t{E}}^{(i+1)} &=  a_{\t{E}}^{(i)}  + \|\alpha\| + 2 \| \beta \| b_{\t{E}}^{(i)} , \\
b_{\t{E}}^{(i+1)} &=  b_{\t{E}}^{(i)} + \|\beta\|,\quad a_{\t{E}}^{(0)} = 0,\quad b_{\t{E}}^{(0)} = 0.
\end{align}
Using the inequality $\| \alpha_{\t{E}}^{(n)} \| \le a_{\t{E}}^{(n)}$, we obtain the upper bound for the QFI associated with (E) strategies
\begin{equation}
F_{\t{E}}^{(n)} \le \min_{\{K_k\}} 4 a_{\t{E}}^{(n)},
\end{equation}
which is equivalent to the state-of-the-art bound presented in \eqref{eq:parbound}.
Again, to make this bound tighter, we can allow for different Kraus representations of an elementary channel in each step of the iteration:
\begin{equation}
\label{eq:D21}
F_{\t{E}}^{(n)} \le \min_{\{K_k\}^{\times n}} 4 a_{\t{E}}^{(n)}.
\end{equation}
From the inequality $\|\beta\|^2 \le \|\alpha\|$ and from the induction principle follows  $b_{\t{E}}^{(i)} \le \sqrt{a_{\t{E}}^{(i)}}$. Therefore, if we compare \eqref{eq:itermin} with \eqref{eq:D18}, we see that $a_{\t{E}}^{(n)} \le a^{(n)}$, so the iterative bound for (E) strategies is tighter than the one for (AD) strategies. However, the minimization from \eqref{eq:D21} cannot be performed step by step as easily as in the adaptive case because the minimimal value of $a_{\t{E}}^{(i+1)}$ is not always achieved for the minimal value of $a_{\t{E}}^{(i)}$.
\section{Asymptotic equivalence between adaptive and parallel strategies}
\label{app:asymp}

For any fixed single-channel Kraus representation $\{K_k\}$ (not necessarily the optimal one), we have:
\begin{equation}
\label{eq:asiter}
a^{(i+1)} =a^{(i)} + \| \alpha\| + 2 \| \beta \|  \sqrt{ a^{(i)}},\quad  a^{(0)}=0.
\end{equation}
We will show that $\forall_{n\geq 1}  a^{(n)}\leq f(n)$, where:
\begin{equation}
f(n)= n \|\alpha\| +  n (n-1)\|\beta\|^2 + n \log n(\|\alpha\|-\|\beta\|^2).
\end{equation}
First note that they are equal at $n=1$:
\begin{equation}
    f(1)= a^{(1)}=\|\alpha\|.
\end{equation}
Next, we will show that this function satisfies the iteration rule \eqref{eq:asiter} with inequality $\geq$ instead of $=$,
\begin{equation}
\label{eq:asstep}
    f(n+1)-f(n)\overset{?}{\geq} \|\alpha\|+2\|\beta\|\sqrt{f(n)}
\end{equation}
which would lead to $f(n)\geq a^{(n)}$. 

To prove \eqref{eq:asstep}, let us write down the l.h.s. explicitly:
\begin{multline}
    f(n+1)-f(n)=\\
    \|\alpha\| + 2 \|\beta\|^2 n + (\|\alpha\| - \|\beta\|^2) ((1 + n) \log(1 + n) -n \log(n)).
\end{multline}
Note that first derivative of $x \log(x)$ is a strictly increasing function, and therefore:
\begin{multline}
\label{eq:E6}
    (1 + n) \log(1 + n) -n \log(n)=\int_n^{n+1}(x\log(x))'dx\geq \\ (x\log(x))'\Big|_{x=n}=1+\log(n).
\end{multline}
Using \eqref{eq:E6}, we can deduce that \eqref{eq:asstep} follows from
\begin{multline}
   \|\alpha\| + 2 \|\beta\|^2 n + (\|\alpha\| - \|\beta\|^2)(1+\log(n))\overset{?}{\geq} \\ \|\alpha\|+2\|\beta\|\sqrt{\|\beta\|^2n(n-1)+n\|\alpha\|+(\|\alpha\|-\|\beta\|^2)n\log(n)},
\end{multline}
which, after subtracting $\|\alpha\|$, taking the square and subtracting the r.h.s. reduces to 
\begin{equation}
    (\|\alpha\| - \|\beta\|^2)^2 (1 + \log(n))^2\geq 0,
\end{equation}
which is always true.

Now, as $f(n)\geq  a^{(n)}$ for any fixed Kraus representation, $4f(n)$ minimized over all Kraus representations constitutes a valid bound for the QFI \eqref{eq:assymptoticbound}.

\section{Continuous time limit for Markovian dynamics}
\label{app:continuous}
We consider a Markovian semigroup dynamics in continuous time, described by the Gorini-Kossakowski-Sudarshan-Lindblad (GKSL) master equation 
\begin{equation}
    \label{eq:GKSL}
    \frac{d \rho }{d t } = - i [ H , \rho ] + \sum_{j=1}^J L_j \rho L_j^\dag - \frac{1}{2} \rho L_j^\dag L_j -  \frac{1}{2} L_j^\dag L_j \rho.
\end{equation}
A Kraus representation of the dynamical map at the lowest order in $\Delta t$ is
\begin{align}
    \label{eq:GKSL_K0}
    K_0 &= \id - \left( \I H + \frac{1}{2} \vect{L}^\dag \vect{L} \right) \Delta t + O\left(\Delta t^2\right) \\
    \vect{K} &= \vect{L} \sqrt{\Delta t} + O\left(\Delta t^{\frac{3}{2}}\right), \label{eq:GKSL_Kvec}
\end{align}
where we introduced a vector notation $\vect{L}=[L_1 , \dots L_J]^T$ for the $J$ collapse operators, which means $ \vect{L}^\dag \vect{L} = \sum_{j=1}^J L_j^\dag L_j$, and for the $J$ Kraus operators $\vect{K} = [ K_1, \dots ,K_J]^T $ (the $0$th operator is kept separate). 
Following Refs.~\cite{Sekatski2016,Demkowicz-Dobrzanski2017}, we obtain the continuous time evolution as the limit $\Delta t \to 0$ of repeated sequential applications of the channel with the above Kraus representation.

Since $F^{(i)} = 4 \| \alpha^{(i)}\|$ when the input state involves ancillary systems~\cite{Fujiwara2008}, we can rewrite Eq.~\eqref{eq:iter} as an inequality for the difference quotient of the (ancilla-assisted) QFI
\begin{equation}
    \label{eq:QFIdiffratiobound}
    \frac{F^{(i+1)} - F^{(i)}}{\Delta t } \leq \frac{4}{\Delta t } \left( \left\Vert  \alpha \right\Vert + \left\Vert \beta  \right\Vert \sqrt{F^{(i)}} \right),
\end{equation}
where $\alpha$ and $\beta$ depend on $\Delta t$ through the Kraus operators in Eqs.~(\ref{eq:GKSL_K0},\ref{eq:GKSL_Kvec}).
 
The parameter $\varphi$ to be estimated is arbitrary and can appear both in the Hamiltonian $H$ and in the collapse operators $L_j$.
The derivatives of the Kraus operators are
\begin{align}
    \dot{K}_0 & = - \left( \I \dot{H} + \frac{1}{2} \dot{\vect{L}}^\dag \vect{L} + \frac{1}{2} \vect{ L}^\dag \dot{\vect{L}} \right) \Delta t \\
    \dot{\vect{K}} & = \dot{\vect{L}} \sqrt{\Delta t}.
\end{align}
The different $\Delta t$-order of the $0$th Kraus operator suggests to divide the matrix $h$ in Eq.~\eqref{eq:KrausTilde} that specifies the Kraus representations in blocks as 
\begin{equation}
    h= \begin{bmatrix}
        h_{00} & \vect{h}^\dag \\
        \vect{h} & \mathfrak{h}
    \end{bmatrix},
\end{equation}
with $h_{00} \in \mathbb{R}$ and $\mathfrak{h}^\dag = \mathfrak{h}$.
We expand this matrix in powers of $\Delta t$ as follows
\begin{equation}
    h = h^{(0)} + h^{(\frac{1}{2})} \sqrt{\Delta t} + h^{(1)} \Delta t,
\end{equation}
and it must be chosen such that the limit for $\Delta t \to 0$ remains finite and Eq.~\eqref{eq:QFIdiffratiobound} gives a meaningful bound on the time-derivative of the QFI.
From Eq.~\eqref{eq:QFIdiffratiobound} it is evident that we need to choose $h$ such that the terms of order less than $\Delta t$ in $\alpha$ and $\beta$ are zero, while the terms of higher order become irrelevant in the limit $\Delta t \to 0$.

After some algebra one realizes that we must impose $h^{(0)}_{00}=h^{(\frac{1}{2})}_{00}=0$ and $\vect{h}^{(0)}=0$ to have $\beta^{(0)}=\beta^{(\frac{1}{2})}=\alpha^{(0)}=\alpha^{(\frac{1}{2})}=0$; the first-order terms that remain relevant in the limit are
\begin{align}
    \nonumber
    -\I \beta^{(1)} =& \dot{H} - \frac{ \I }{2} \left( \dot{\vect{L}}^\dag \vect{L} - \vect{L}^\dag \dot{\vect{L}} \right)  + h_{00}^{(1)} \id + \vect{L}^\dag \vect{h}^{(\frac{1}{2})} + \vect{h}^{(\frac{1}{2})\dag} \vect{L} \\
    & + \vect{L}^\dag \mathfrak{H}^{(0)} \vect{L}
\end{align}
and
\begin{equation}
    \alpha^{(1)} = \left[ \vect{h}^{(\frac{1}{2})} \id  + \mathfrak{h}^{(0)} \vect{L} + \I \dot{\vect{L}} \right]^\dag \left[\vect{h}^{(\frac{1}{2})} \id  + \mathfrak{h}^{(0)} \vect{L} + \I \dot{\vect{L}}   \right].
\end{equation}
The bound on the rate of increase of the QFI is
\begin{equation}
    \label{eq:continuousQFIderivative}
        \frac{d F(t)  }{ d t } \leq 4 \min_{h_{00}^{(1)}, \vec{h}^{(\frac{1}{2})},\mathfrak{h}^{(0)} } \left( \left\Vert \alpha^{(1)} \right \Vert +  \left\Vert \beta^{(1)} \right \Vert \sqrt{ F(t) }   \right).
\end{equation}
Since Eq.~\eqref{eq:iter} can also be derived when a different channel is applied at each step, the previous inequality also holds in the Markovian time-inhomogenous case with time-dependent $H$ and $L_j$ in Eq.~\eqref{eq:GKSL}.

Recently, the authors of Ref.~\cite{Wan2022} derived a bound tighter than the one in Refs.~\cite{Demkowicz-Dobrzanski2017,Zhou2017}, without relying on a discretization of the time evolution.
For a Hamiltonian parameter (i.e. $\dot{\vect{L}}=0$), Eq.~(18) of Ref.~\cite{Wan2022} gives a state-dependent bound for arbitrary initial states.
The inequalities in Eq.~(19) of Ref.~\cite{Wan2022} can further be applied to obtain a state-independent bound for finite-dimensional systems which coincides with the bound in Eq.~\eqref{eq:continuousQFIderivative} by applying the substitutions $-\I \beta^{(1)} \leftrightarrow G$, $\alpha^{(1)} \leftrightarrow \sum_j A_j^\dag A_j $, $ h_{00}^{(1)} \leftrightarrow - x$, $h^{(\frac{1}{2})}_j \leftrightarrow - \beta_j$ and $\mathfrak{h}^{(0)}_{ij} \leftrightarrow - \gamma_{ij}$. 

For a Hamiltonian parameter, the two asymptotic cases are
\begin{enumerate} 
    \item The Hamiltonian derivative $\dot{H}$ is in the Lindblad span and we can find $h$ such that $\beta^{(1)} = 0$ leading to 
    \begin{equation} F \leq  4 t \min_{h_{00}^{(1)}, \vec{h}^{(\frac{1}{2})},\mathfrak{h}^{(0)}, \beta^{(1)}=0 } \left\Vert \alpha^{(1)} \right\Vert
    \end{equation}
    \item The Hamiltonian derivative $\dot{H}$ is not in the Lindblad span, then the asymptotic solution of the differential equation for large $t$ gives
    \begin{equation}
        F \leq  4 t^2 \min_{h_{00}^{(1)}, \vec{h}^{(\frac{1}{2})},\mathfrak{h}^{(0)} } \left\Vert \beta^{(1)} \right\Vert^2.
    \end{equation}
\end{enumerate}
Both bounds are asymptotically attainable with the parallel strategies described in Refs.~\cite{Zhou2017,Zhou2019e}.
However, Eq.~\eqref{eq:continuousQFIderivative} can give tighter bounds for short times~\cite{Wan2022}.
Finally, while the estimation of a parameter appearing in the collapse operators is less studied, the bounds of Refs.~\cite{Demkowicz-Dobrzanski2017,Zhou2017} have recently been applied to study optimal quantum thermometry in Markovian environments~\cite{Sekatski2021}, thus these tighter bounds may also be useful in that context.

\end{document}